\documentclass[aip,jcp,floatfix,reprint]{revtex4-1}
\usepackage{tabularx}
\newcolumntype{Y}{>{\centering\arraybackslash}X}
\usepackage{color} 
\usepackage{graphicx} 
\usepackage{amsmath, amsthm, amssymb}
\usepackage{amsthm}
\usepackage{multirow}
\usepackage{ulem} 
\begin{document} 
\title{Effect of simple solutes on the long range dipolar correlations in liquid water} 
\author{Upayan Baul} 
\email{upayanb@imsc.res.in} 
\affiliation{The Institute of Mathematical Sciences, C.I.T. Campus, Taramani, Chennai 600113, India} 
\author{J. Maruthi Pradeep Kanth}
\email{jmpkanth@imsc.res.in}
\affiliation{Vectra LLC, Mount Road, Chennai - 600006, India}
\author{Ramesh Anishetty} 
\email{ramesha@imsc.res.in} 
\affiliation{The Institute of Mathematical Sciences, C.I.T. Campus, Taramani, Chennai 600113, India} 
\author{Satyavani Vemparala} 
\email{vani@imsc.res.in} 
\affiliation{The Institute of Mathematical Sciences, C.I.T. Campus, Taramani, Chennai 600113, India}
\date{\today}

\begin{abstract} 
Intermolecular correlations in liquid water at ambient conditions have generally been characterized through short range density fluctuations described 
through the atomic pair distribution functions (PDF). Recent numerical and experimental results have suggested that such a description of order or 
structure in liquid water is incomplete and there exists considerably longer ranged orientational correlations in water that can be studied through dipolar 
correlations. In this study, using large scale classical, atomistic molecular dynamics (MD) simulations using TIP4P-Ew and TIP3P models of water, we show that 
salts such as sodium chloride (NaCl), potassium chloride (KCl), caesium chloride (CsCl) and magnesium chloride (MgCl$_2$) have a long range effect on 
the dipolar correlations, which can not be explained by the notion of structure making and breaking by dissolved ions. The relative effects of cations 
on dipolar correlations are observed to be consistent with the well-known Hofmeister series. Observed effects are explained through orientational stratification 
of water molecules around ions, and their long range coupling to the global hydrogen bond network by virtue of the sum rule for water. The observations 
for single hydrophilic solutes are contrasted with the same for a single methane (CH$_4$) molecule. We observe that even a single small hydrophobe can result 
in enhancement of long range orientational correlations in liquid water,- contrary to the case of dissolved ions, which have been observed to have a reducing effect. 
The observations from this study are discussed in the context of hydrophobic effect.
\end{abstract}

\maketitle

\section{Introduction}
\indent
Liquid water and aqueous solutions, ubiquitous in nature, provide the native environment for numerous processes taking place in nature. Detailed 
understanding of structural and dynamical behavior of water and the effect of simple salts (or ion combinations) on the same are thus critical 
to the understanding of further complex and interesting processes of biological~\cite{kunz04,nostro,ball15} and environmental~\cite{buszek,jung06} 
relevance. Many of the unique properties of water can be attributed to the ability of water molecules to form an extensively connected network of 
hydrogen bonds in which the four nearest neighbors of a water molecule arrange themselves in a nearly tetrahedral geometry~\cite{eisen69,ziel05,stan00,kumar09}. 
This local structuring, in addition to the generally observed short range structure in liquids, owing to the excluded volume effects, has led to 
consensus in treating water as a highly structured liquid~\cite{marcus09chemrev,str1,str2,str3,str4,str5}. However, the precise definition of 
microscopic structure in water is not without ambiguity~\cite{marcus09chemrev,ball15}. 

\indent
The structure of liquid water has been extensively probed using pair distribution functions (PDFs) among its atomic constituents using experimental as 
well as numerical techniques both in its bulk liquid phase and in presence of ionic impurities~\cite{gordon02,marcus09chemrev}. Observed PDFs 
have suggested~\cite{t4pew,5p,t3p, gr1,kanth10} that at ambient conditions, the structure of liquid water is short range and the spatial distribution of water molecules are 
uncorrelated beyond a distance of $\sim(8-10)\mathring{A}$. However, PDFs yield a measure for density fluctuations 
alone and are devoid of information on orientational correlation, especially at longer separations. The mutual orientations of water molecules separated 
in space can, however, be envisaged through correlations involving dipolar and higher order multipole moments of the constituent molecules~\cite{mat04,kanth10,dipzhang14,elton14}. 
Interestingly, while quadrupolar fluctuations have been observed to disappear within a distance of $\sim3\mathring{A}$, correlations involving the 
dipolar degree of freedom of water molecules have been observed to be considerably longer ranged~\cite{kanth10}. These dipole - dipole correlations 
were decomposed in the same report into transverse or the trace part, involving the alignment of the dipoles with respect to themselves and longitudinal 
or the traceless part, measuring the alignment of the dipoles with respect to the radial vector separating them~\cite{kanth10}. Both of these 
correlations show oscillatory solvation structure and are longer ranged compared to density 
correlations in water. The former vanishes beyond $\sim14\mathring{A}$ in liquid water in compliance with rotational symmetry. The latter, 
oscillatory in nature and always positive, was observed~\cite{kanth10} to be non-vanishing even at $\sim75\mathring{A}$ separations and decays exponentially beyond solvation 
region ($14\mathring{A}$) with largest correlation length of $\sim24\mathring{A}$. A non-vanishing alignment of 
dipole vectors with respect to each other over separations comparable to system dimensions would indicate the existence of spontaneous polarization 
in the medium. These results have been observed to be in good agreement with further numerical observations of long range angular correlations in liquid 
water~\cite{angliu13,dipzhang14}. Another dipolar correlation of interest for bulk liquid water is 
the oxygen - dipole correlation which represents the propensity for alignment of a dipole vector to the radially outward direction when observed from the 
position of another water molecule. This correlation also shows oscillatory solvation structure at radial separations below $14\mathring{A}$ and vanishes 
beyond the same~\cite{kanth10}.

\indent
The experimental observations of such long range dipolar or angular correlation are not accessible to conventional spectroscopic techniques owing to the 
lack of positional ordering at longer length scales. However, recent hyper-Rayleigh light scattering (HRS) experiments~\cite{hrs1,hrs2,hrs3,hrs5} have 
revealed the existence of long range orientational correlations longer than 230$\mathring{A}$ in liquid water~\cite{angliu13}. Dipolar correlations in 
liquid water have been studied for a variety of non-polarizable as well as polarizable models of water, including (TIP5P, TIP3P)~\cite{kanth10} and 
(SPC/E, swm4-DP)~\cite{dipzhang14}. Within the limitations of system sizes studied, the results for the various water models have been consistent. 
With the SPC/E and swm4-DP water models, the longitudinal part of the correlations was qualitatively observed to be in agreement with the results for 
TIP5P and TIP3P water models~\cite{dipzhang14,kanth10} for radial separations $\leq12\mathring{A}$. The longer correlation length of $\sim24\mathring{A}$, 
however, requires considerably larger system sizes than reported~\cite{dipzhang14}. In a recent work~\cite{hrs5}, correlation functions computed 
from HRS measurements were directly compared with the longitudinal and transverse correlations from MD simulations~\cite{dipzhang14}. While the experimentally 
observed correlations did not reproduce the oscillatory solvation structure, the asymptotic behavior was observed to be in good qualitative agreement. 
The long range correlations from HRS experiments were stated to have $r^{-3}$ dependence, which indicative of the importance of dipole - dipole interactions.

\indent
It is important to note that the long range behavior of the longitudinal correlation is not a consequence of long range electrostatic interactions, but has 
its origin in the fluctuations of the underlying hydrogen bond network of water. The same was established in our previous work, where 
truncating the electrostatic interactions beyond $12\mathring{A}$ was observed to have no effect on the asymptotic behavior of the longitudinal 
correlation~\cite{kanth10}. While hydrogen bonding interactions are intrinsically short range, the fluctuations of the hydrogen bond network are 
restricted by topological constraints. The entropy of hydrogen bond network thus has contributions from both single-water and collective 
dynamics~\cite{netz}. In a statistical description of liquid water, the hydrogen bond fluctuations need to be consistent with a \textquotedblleft sum rule\textquotedblright, 
which states that sum of density of dangling bonds (i.e., a water molecule\textquoteright s bond arms which are not hydrogen bonded to any other molecule) and twice the 
density of hydrogen bonds should be equal to four times the density of water molecules~\cite{kanth12}. The dynamics of the hydrogen bond network 
of water, governed by orientational fluctuations of water molecules, thus results in orientational correlations that are substantially long 
range compared to the length scale ($\sim 2 \mathring{A}$) of hydrogen bonding interactions~\cite{kanth10,kanth12,kanth13}.

\indent
Temperature and density have been observed to have small but monotonic effects on the transverse and longitudinal dipolar correlations~\cite{kanth10,dipzhang14}. 
Presence of salt (CaCl$_2$), however, was observed to have a considerably larger effect at reported~\cite{dipzhang14} length scales of $\leq15\mathring{A}$. 
The system sizes reported therein, with $\sim432$ water molecules for CaCl$_2$ - water systems, are however, unlikely to capture the quantitative aspects of 
the effects of salts on the long range correlations. 

\indent
Over decades, effort has been made to characterize the effect of salts on the structure of water based on the ability of the dissolved ions to enhance or reduce 
the same~\cite{marcus09chemrev}. Consequently, ions have often been characterized as \textit{structure makers} and \textit{structure breakers}. This putative notion 
has been recently subjected to considerable criticism, primarily owing to the absence of an unambiguous definition of structure in water~\cite{ball15,makbr1,makbr2}. 

\indent
In the current study, we investigate the effect of multiple salts (CsCl, KCl, NaCl and MgCl$_2$) on the dipolar correlations for large system sizes using TIP4P-Ew and 
TIP3P water models. Our results clearly demonstrate that structure making (and breaking) is not a generic concept, but crucially depends on the particular correlation 
in water. Without invoking a definition of structure, we discuss the consistency of our observations in a general framework motivated by the sum rule and associated 
orientational entropy. For the same, we invoke electrostatics driven orientational stratification of water molecules situated close to the ions. To quantify the same, 
we define a solute - dipole correlation which is a trivial extension of the oxygen - dipole correlation. Finally, as a first extension to the case of non - 
ionic solutes, we contrast the effects observed for single salt molecules dissolved in water with the same for a methane molecule and show that hydrophilic and hydrophobic 
impurities of comparable sizes differ characteristically in their effects on the long range orientational correlations in water. We discuss the implications of this observation in the context of hydrophobic 
effect.

\indent
The rest of the article is arranged as follows: in section II, we define the dipolar correlations used in the current study (elaborated in reference~\cite{kanth10}) 
followed by a description of systems studied and simulation details. The results obtained are discussed in section III, followed by discussions and future directions 
in section IV.

\section{Methods}
\subsection{Dipolar correlations}
\subsubsection{dipole - dipole correlations}
\indent
Let $\hat{\mu}$ denote the normalized dipole vector of a water molecule. Statistical correlations among any two water molecules, whose oxygen atoms 
are situated at $\bf{r_1}$ and $\bf{r_2}$, involving the dipolar degree of freedom can be formulated~\cite{kanth10} as 
\begin{equation}
 \langle \hat{\mu}^i (\mathbf{r_1}) \hat{\mu}^j (\mathbf{r_2}) \rangle = \frac{1}{2} \left( \delta^{ij} - \frac{r^i r^j}{r^2} \right)t(r) - \frac{1}{2} \left( \delta^{ij} - 3\frac{r^i r^j}{r^2} \right)l(r)
\end{equation}
where indices $i,j$ denote directions in three-dimensional space, $\mathbf{r}=(\mathbf{r_1} - \mathbf{r_2})$, $r=|\mathbf{r}|$ and the angular 
brackets denote ensemble averages. The decomposed scalar functions $t(r)$ and $l(r)$, defined as 
\begin{eqnarray}
 t(r) & \,=\, & \langle \hat{\mu}(\mathbf{r_1}) . \hat{\mu}(\mathbf{r_2}) \rangle \\
 l(r) & \,=\, & \langle \hat{\mu}(\mathbf{r_1}) . \hat{\mathbf{r}} \, \hat{\mu}(\mathbf{r_2}) . \hat{\mathbf{r}} \rangle
\end{eqnarray}
describe the transverse (trace) part and the longitudinal (traceless) part of the tensorial correlations $\langle \hat{\mu}^i (\mathbf{r_1}) \hat{\mu}^j (\mathbf{r_2}) \rangle$. 
Physically, $t(r)$ and $l(r)$ are measures for the statistical alignment of water dipole vectors spaced $r$ distance apart with respect 
to themselves and with respect to the radial vector separating them respectively.

\subsubsection{position - dipole correlations}
\indent
The oxygen - dipole correlation ($d(r)$) is defined as~\cite{kanth10}
\begin{equation}
 d(r)\,=\,\langle \hat{\mu}(\mathbf{r_1}) \ldotp \mathbf{\hat{r}} \rho (\mathbf{r_2})\rangle
\end{equation}
where $\rho(\mathbf{r})$ takes a value of 1 or 0 depending on presence or absence of oxygen atom at $\mathbf{r}$. Similarly, the solute - dipole correlation can be defined 
through the same expression, where the density field for oxygen atoms ($\rho$(${\mathbf{r}}$)) is replaced by the same for a solute species. The oxygen - dipole and solute - 
dipole correlations represent the propensity for alignment of the dipole vector of a water molecule to the radially outward direction, when observed from the position of 
another water molecule or a solute respectively.

\subsection{System setup}
\indent
Classical atomistic molecular dynamics simulations were performed using simulation package NAMD 2.9~\cite{namd} using TIP4P-Ew~\cite{t4pew} and TIP3P~\cite{t3p} water models. 
Recently developed parameters that correctly reproduce the right coordination 
number were used for simulating divalent Mg$^{2+}$ ions with TIP4P-Ew water model~\cite{divt4p}. Rest of the monovalent ion parameters 
for simulating TIP4P-Ew systems were taken from extensively used halide and alkali ion parameters available in literature~\cite{monot4p}. For simulating 
aqueous solutions using TIP3P water model, standard CHARMM parameters~\cite{iont3p}, used extensively in biomolecular systems, were used for all ions. Methane parameters were 
derived from aliphatic \textit{sp3} carbon and nonpolar hydrogen parameters in CHARMM~\cite{charmmProt}.

\begin{figure*}[ht]
 \centering
 \includegraphics[scale=1.4]{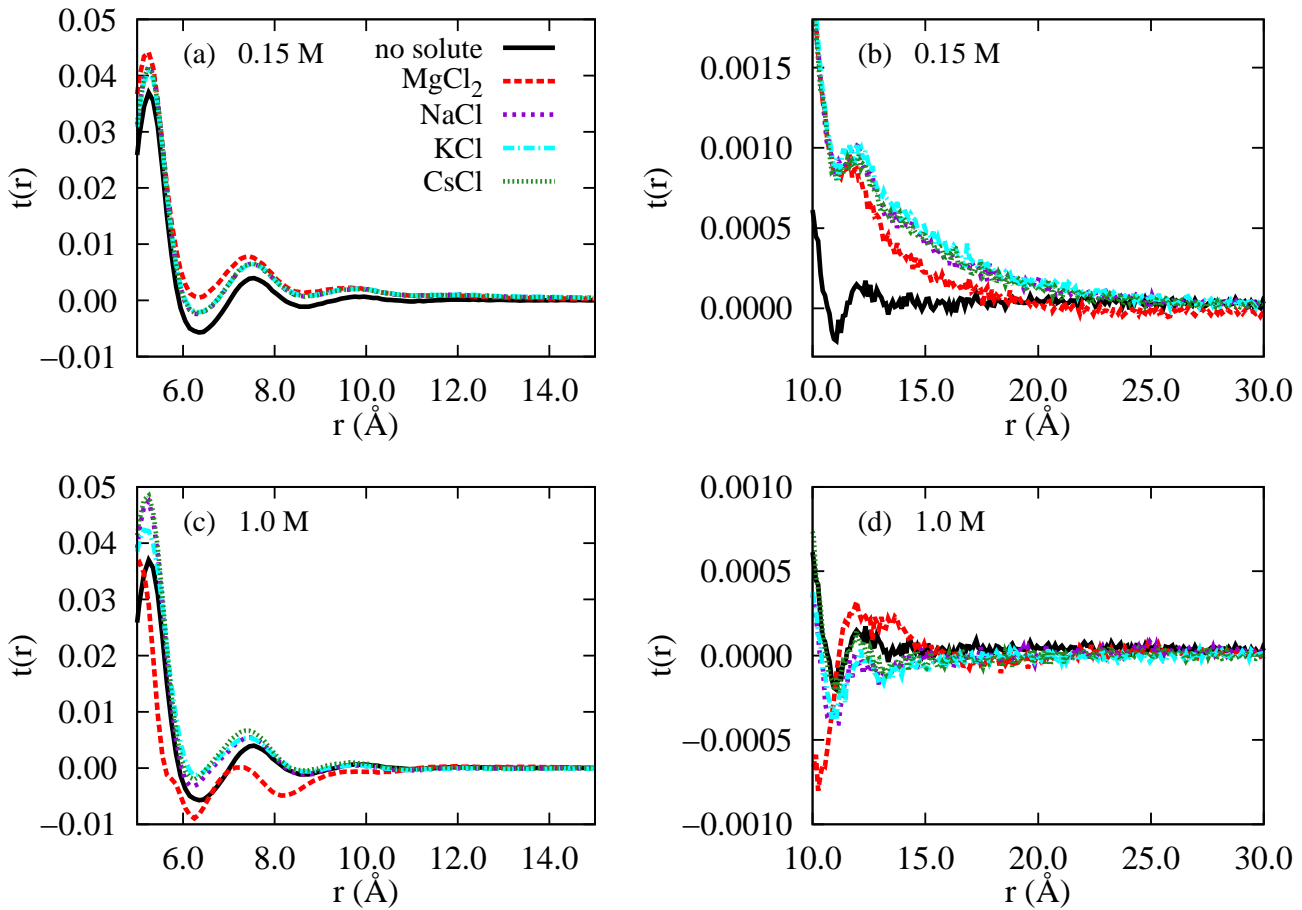}
 \caption{\small (color online) $t(r)$ correlation functions for systems with TIP4P-Ew water model at 0.15M and 1.0M salt concentrations. In all plots, the curve for 
 water in absence of solutes is shown in (solid, black) line. The same are shown in (dashed, red), (dotted, purple), (dash-dot, cyan) and (small dots,  green) for MgCl$_2$, 
 NaCl, KCl and CsCl solutions respectively. Error bars have not been plotted since the errors have been observed to be small compared to the mean values, especially at 
 larger separations (see \textit{Supp. Info.}~\cite{supplement} for explanation).}
\end{figure*}

\begin{figure*}[ht]
 \centering
 \includegraphics[scale=1.4]{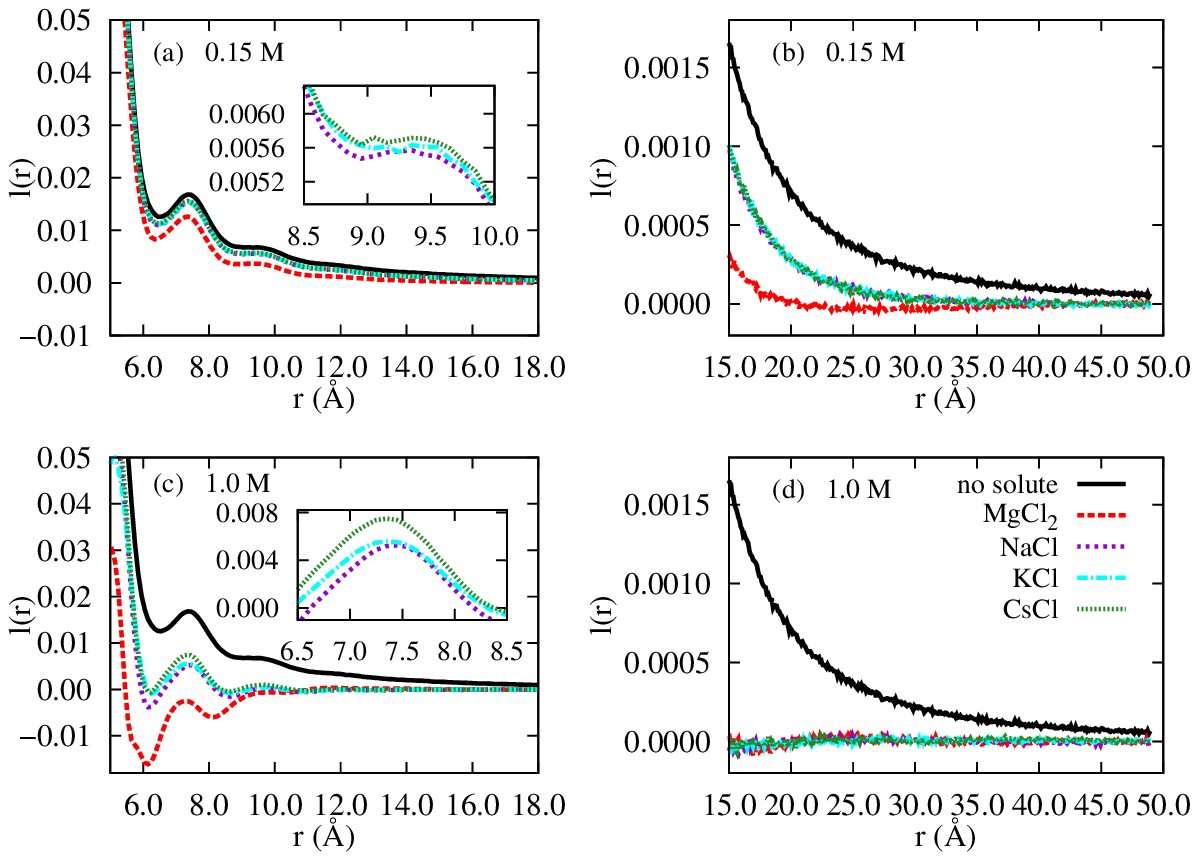}
 \caption{\small (color online) $l(r)$ correlation functions for systems with TIP4P-Ew water model at 0.15M and 1.0M salt concentrations. In all plots, the curve for 
 water in absence of solutes is shown in (solid, black) line. The same are shown in (dashed, red), (dotted, purple), (dash-dot, cyan) and (small dots, green) for MgCl$_2$, 
 NaCl, KCl and CsCl solutions respectively. Error bars have not been plotted since the errors have been observed to be small compared to the mean values, especially at 
 larger separations (see \textit{Supp. Info.}~\cite{supplement} for explanation).}
\end{figure*}
\indent
Given the longest correlation length is $\sim24\mathring{A}$ for $l(r)$ dipolar correlations~\cite{kanth10}, cubic box lengths of $\gtrsim50\mathring{A}$ 
need to be simulated for quantitative estimates of the same. As to enable the identification of any enhancement of the same, or the appearances 
of even longer correlation lengths, if any, considerably larger cubic boxes of dimensions $\approx 100\mathring{A}\times100\mathring{A}\times100\mathring{A}$ 
were simulated in the investigations of $t(r)$ and $l(r)$ correlations. Solute free water systems comprising of 36054 water molecules for TIP4P-Ew and 34194 water 
molecules for TIP3P were initially equilibrated for 5ns under isothermal, isobaric (NPT) conditions under a pressure of 1atm. The temperature was maintained at 298K for TIP4P-Ew 
and at 305K for TIP3P. The system configurations at the end of the NPT equilibration were used to generate 0.15M and 1.0M aqueous solutions of CsCl, NaCl, 
KCl and MgCl$_2$ using \textit{solvate} plugin of VMD~\cite{vmd}. The salt-water systems were further equilibrated for 5ns under NPT 
conditions (pressure 1atm; temperature 298K and 305K for TIP4P-Ew and TIP3P respectively). All systems, including solute free water systems 
were equilibrated for further 5ns under NVE conditions resulting in total equilibration time of 10ns for each system. Following equilibration, 
production runs for all systems were carried out for additional 10ns, over which the system configurations were written every 4ps. The analyses 
for each system were performed on the 2500 uncorrelated system configurations thus generated. 

\indent
For comparative study of effects of single hydrophilic versus hydrophobic solutes, and the analyses of $d(r)$ correlations, a pre-equilibrated water box of 
dimensions $\approx 50\mathring{A}\times50\mathring{A}\times50\mathring{A}$ was extracted from the 10ns equilibrated configuration of the 
$\approx 100\mathring{A}\times100\mathring{A}\times100\mathring{A}$ water box for TIP4P-Ew water model. Single MgCl$_2$, NaCl and CH$_4$ molecules were added to the same to 
construct extremely dilute (1 solute molecule in $\sim$4100 water molecules) solutions. All four systems were further equilibrated for 2ns under NPT conditions (pressure 1atm; 
temperature 298K) followed by 3ns under NVE. Production runs for 10ns were then carried out for each of the three systems under NVE conditions, generating 5000 configurations for analyses. 
Ion pairing effects were not observed for the salt solutions studied here.

\indent
For the NPT equilibration stages, constant pressure and temperature were maintained using Langevin Piston~\cite{pres} and temperature coupling to external 
reservoir respectively. Timesteps of 1fs and 2fs were used for simulating systems with TIP4P-Ew and TIP3P water model respectively. For all simulations, long 
range electrostatic interactions were computed using particle mesh Ewald (PME) and Lennard-Jones interactions were smoothly truncated 
beyond $12\mathring{A}$ through the use of switching functions between $10\mathring{A}$ and $12\mathring{A}$.

\section{Results}
\indent
In the following subsections we report the observations of the effects of studied solutes on the dipolar correlations $t(r)$, $l(r)$ and $d(r)$. All the results in 
the following subsections are for TIP4P-Ew water model and the corresponding results for TIP3P water model are given in \textit{Supp. Info}~\cite{supplement}. 
In subsections A and B we describe the effects of salts on the $t(r)$ and $l(r)$ correlations at salt concentrations of 0.15M and 1.0M. In subsection 
C we describe $d(r)$ correlation, both in absence of salts and around cationic species. We conclude by comparing results for the presence of single molecules of NaCl, 
MgCl$_2$ and CH$_4$ in subsection D.

\subsection{Effect of salts on transverse correlations}
\indent
Computed $t(r)$ correlation functions for systems with TIP4P-Ew water model are shown in FIG. 1 (see FIG. SI of \textit{Supp. Info}~\cite{supplement} 
for similar plots with TIP3P water model). For clarity of comparison, the plots have been shown for 
$r \geq 5\mathring{A}$. Plots for the full range of $r$ are shown in FIG. SIV of \textit{Supp. Info}~\cite{supplement}. In good agreement with prior results for solute free 
liquid water~\cite{kanth10,dipzhang14}, $t(r)$ shows oscillatory solvation structure over radial separations below $14\mathring{A}$ and vanishes beyond the same. 
All salts at low concentration (0.15M) enhance $t(r)$ in comparison with solute free water at both smaller ($\leq10\mathring{A}$) and larger ($\geq10\mathring{A}$) 
separations (FIG. 1(a,b)). At smaller separations, the enhancement is observed to be most pronounced in the presence of MgCl$_2$. The other salts studied (NaCl, KCl and CsCl) are observed 
to induce smaller, roughly equal enhancements in $t(r)$. Beyond $10\mathring{A}$, however, the effects are reversed with NaCl, KCl and CsCl resulting in a greater enhancement in $t(r)$ over 
MgCl$_2$. Further, all salts at 0.15M concentration can clearly be observed to lead to an enhancement in the range of $t(r)$, being non-vanishing till $\sim 18 \mathring{A}$ 
for MgCl$_2$ and $\sim 24 \mathring{A}$ for the rest. No appreciable shift in peak positions is observed at 0.15M concentration for all salts.

\indent
At 1.0M concentration , the effect of MgCl$_2$ can be observed to be substantially different from that of the other salts, as can be seen in FIG. 1(c). While NaCl, 
KCl and CsCl still result in an enhancement of $t(r)$ over that for solute free water for smaller separations, MgCl$_2$ leads to a reduction of the 
same beyond the first solvation shell. The effect is most prominently observable beyond the second solvation peak, where presence of MgCl$_2$ 
results in an anticorrelation in the transverse part of dipolar correlations. Effects of NaCl, KCl and CsCl are also distinguishable at the 
higher concentration, with CsCl resulting in a greater enhancement in peak heights for $t(r)$ over NaCl and KCl. The positions of the second 
and third solvation peaks are visibly shifted to left for all salts, the shift being maximum for MgCl$_2$. Further, the appearance 
of a quasi long range nature in $t(r)$ at the lower concentration (0.15M) is washed off at the higher (1.0M) (see FIG. 1(b,d)). The concentration dependent 
enhancing / reducing effect observed for MgCl$_2$ is in good agreement with reported results~\cite{dipzhang14} for CaCl$_2$ which was also observed to affect enhancement 
in $t(r)$ at 0.25M concentration followed by similar anticorrelations at the concentration of 1.56M. The trait thus appears to be a 
general characteristics for salts with strongly solvated cations capable of inducing strong perturbations in the orientation of water molecules.

\subsection{Effect of salts on longitudinal correlations}
\indent
Longitudinal correlations $l(r)$ are of greater interest over $t(r)$ in the study of long range dipolar correlations in water since $l(r)$ 
can be described as a truly long range correlation with an exponential decay~\cite{kanth10}. In solute free liquid water at ambient conditions, $l(r)$ has been 
shown~\cite{kanth10} to exhibit solvation peaks till $14 \mathring{A}$, beyond which it decays exponentially with longest correlation length 
of $\sim 24 \mathring{A}$ and is non-vanishing even at $75 \mathring{A}$. It is always positive in solute free water and decays in an 
oscillatory manner. The characteristics of this function are closely shared by another long range angular correlation function computable 
using classical density functional theory calculations~\cite{angliu13}, further strengthening the likely existence of long range orientationally 
correlated domains of water molecules. It has been suggested from HRS observations that long range orientational correlations in 
molecular liquids are expressed through propagating waves in molecular reorientation instead of diffusional orientation of molecules~\cite{hrs4}. 
$l(r)$ as well as the stated angular correlation function~\cite{angliu13} are thus in qualitative agreement with HRS results. Recently, correlations 
from HRS measurements have been directly compared to $l(r)$ and a linear combination of it with $t(r)$, computed using MD simulations~\cite{hrs5}. 
Results have been in qualitative agreement with similar asymptotic behavior. However, the oscillatory nature of the correlations computed using 
molecular simulations have not been reflected in the HRS results. A direct quantitative comparison with the experimental results is still not possible 
since (a) system dimensions studied using HRS are well beyond the scope of atomistic simulations~\cite{hrs3} and (b) HRS does not provide explicit 
information of microscopic correlations among molecular dipoles~\cite{angliu13}.
\begin{table*}[ht] 
\begin{center}
\begin{tabularx}{\textwidth}{@{}|Y|Y|Y|Y|Y|Y|@{}}
 \hline
 water model & salt (concentration) & $a_1$ & $r_1$ & $a_2$ & $r_2$  \\
 \hline
 \multirow{3}{*}{TIP4P-Ew} & none & 0.31 ($\pm$ 0.01) & 4.82 ($\pm$ 0.08) & 0.0207 ($\pm$ 0.0008) & 24.0 ($\pm$ 0.6)  \\
    & CsCl (0.15M) & 0.321 ($\pm$ 0.004) & 4.93 ($\pm$ 0.02) &   &    \\
    & KCl (O.15 M) & 0.34 ($\pm$ 0.01) & 4.82 ($\pm$ 0.03) &   &    \\
    & NaCl (O.15 M) & 0.34 ($\pm$ 0.01) & 4.77 ($\pm$ 0.03) &   &    \\
    & MgCl$_2$ (O.15 M) & 0.74 ($\pm$ 0.04) & 2.94 ($\pm$ 0.04) &   &   \\
 \hline
 \multirow{3}{*}{TIP3P} & none & 0.34 ($\pm$ 0.01) & 5.24 ($\pm$ 0.12) & 0.026 ($\pm$ 0.001) & 24.8 ($\pm$ 0.9) \\
    & CsCl (0.15M) & 0.344 ($\pm$ 0.006) & 5.95 ($\pm$ 0.04) &   &    \\
    & KCl (O.15 M) & 0.359 ($\pm$ 0.006) & 5.93 ($\pm$ 0.04) &   &    \\
    & NaCl (O.15 M) & 0.323 ($\pm$ 0.005) & 6.10 ($\pm$ 0.04) &   &    \\
    & MgCl$_2$ (O.15 M) & 0.82 ($\pm$ 0.04) & 3.01 ($\pm$ 0.03) &   &   \\
 \hline
 \end{tabularx}
\end{center}
\caption{Numerical fitting results for $l(r)$ for TIP4P-Ew and TIP3P water model systems. For the higher (1.0M) salt concentration, all 
salts result in disappearance of the long range component of $l(r)$. The numbers within parenthesis represent the numerical fitting 
error estimates. The results for TIP4P-Ew and TIP3P water models are at 298K and 305K respectively.}
\end{table*}
\begin{figure*}[ht]
 \centering
 \includegraphics[scale=1.4]{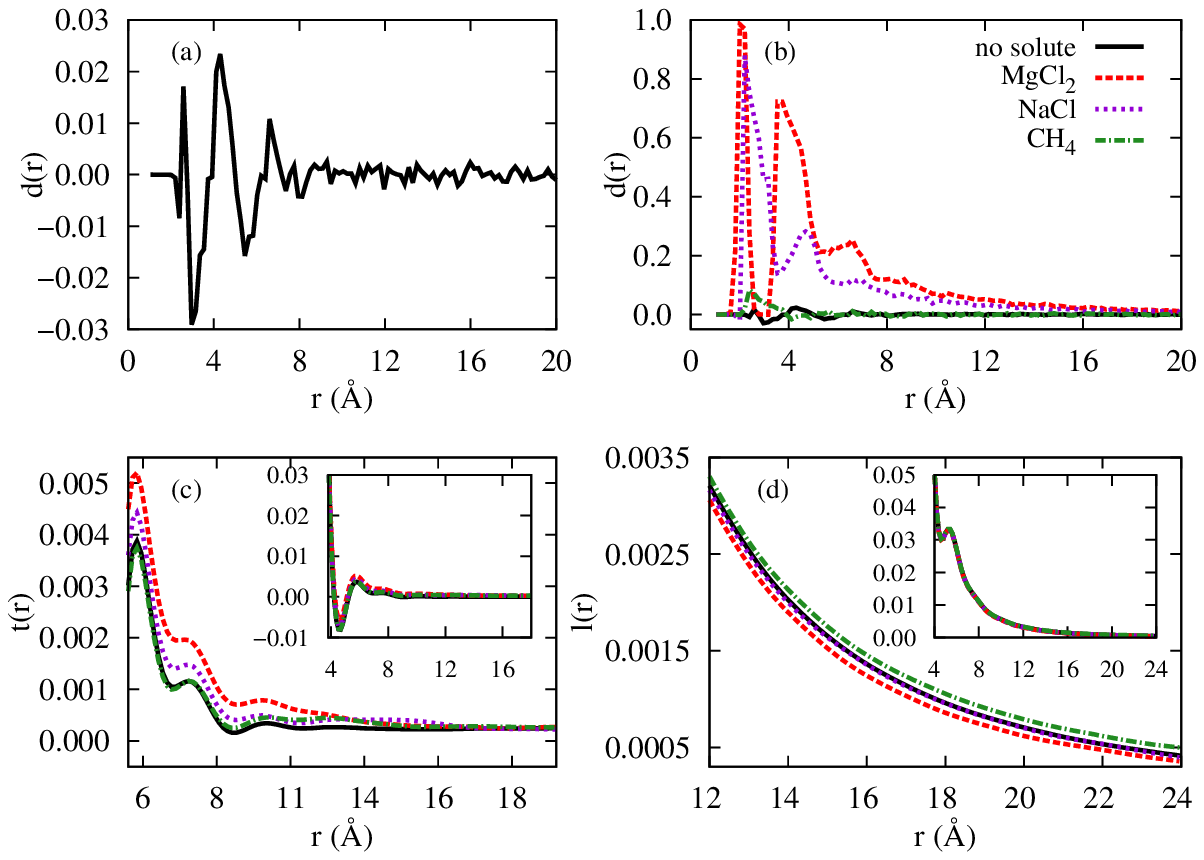}
 \caption{(color online) \textbf{(a)} oxygen - dipole correlation for solute free water. \textbf{(b)} Comparison of $d(r)$ (oxygen - dipole / solute - dipole) correlations as seen by a 
 water molecule in absence of solutes, a methane molecule and a lone cation (with counterion(s)) in a water-box. Curves for solute free water are plotted using (solid, black) lines. 
 The same for MgCl$_2$, NaCl and CH$_4$ are plotted using (large dots, red), (small dots, purple) and (dash-dots, green) respectively.}
\end{figure*}

\indent
Computed $l(r)$ correlations for systems with TIP4P-Ew water model are shown in FIG. 2 and the same for systems with TIP3P model of water are 
included in the \textit{Supp. Info}~\cite{supplement} (FIG. SII). As with $t(r)$, the plots in the article have been shown for 
$r \geq 5\mathring{A}$. Plots for the full range of $r$ are shown in FIG. SIV of \textit{Supp. Info}~\cite{supplement}. 
Given the system sizes ($\sim 100 \times 100 \times 100 \mathring{A}^3$) studied in the current work, the 
correlations have been computed upto $48 \mathring{A}$ separations. The results for solute free water has been observed to be consistent with prior 
results~\cite{kanth10,dipzhang14}. All salts have been observed to induce reduction in both the strength and range of $l(r)$ correlations at both 
the concentrations studied. The reduction is more prominent at the higher salt concentration (1.0M) studied. At 0.15M salt concentrations, the 
$l(r)$ correlations for salt - water systems retain the essential solvation structures of solute free $l(r)$, such as the positions of solvation 
peaks and the absence of anticorrelation. However, the long range the decay of the correlations, which are bi-exponential for solute free water, 
are severely effected even at 0.15M salt concentration (FIG. 2(a,b)). The longest correlation length $\sim 24 \mathring{A}$ for solute free water is not observed 
for any of the salts, even at 0.15M concentrations. The correlation lengths have been obtained by fitting $l(r)$ for $r > 12\mathring{A}$ to 
a bi-exponential or a mono-exponential ($a_2 \,=\,0$, for salt - water systems) function of Ornstein-Zernike kind~\cite{OZmarch}. 
\begin{equation}
l(r) = \frac{a_{1}}{r}\exp(-r/r_{1}) + \frac{a_{2}}{r}\exp(-r/r_{2})
\end{equation}
The fit results are shown in TABLE I for both TIP4P-Ew and TIP3P water models. Within the limitations of model dependence, the results from both 
water models are in good agreement. The comparable effects of NaCl, KCl and CsCl are reflected in the correlation lengths too.

\indent
While the correlation lengths of $l(r)$, computed for individual salt - water systems, can not distinguish the relative effects of NaCl, KCl and CsCl, 
their relative influences are discernible even at 0.15M concentrations through their solvation peaks (as shown in the inset of FIG. 2(a) 
for the third solvation peak). Owing to the common choice of anion (Cl$^-$), the cations studied can be ordered based on their relative effect 
on the longitudinal correlation $l(r)$. The order, in terms of increasing effect on $l(r)$ is observed to be Cs$^+ <$K$^+ <$Na$^+ <$Mg$^{2+}$ 
which is consistent with the ordering of the same ions in the cationic Hofmeister series~\cite{tiel}. The same trend is also observed to be retained at 
1.0M salt concentrations.

\indent
At the higher concentration of 1.0M, salts are observed to severely reduce the $l(r)$ correlations. The absence of anticorrelations in $l(r)$ is observably lost 
for MgCl$_2$ and NaCl as seen from FIG. 2(c). Closer scrutiny of curves for KCl and CsCl also indicate anticorrelated regions. The trends ensure 
that at higher concentrations, all studied salts would result in negatively correlated regions in $l(r)$ profile. Salts studied have not been 
observed to have any appreciable effect on the positions of observable solvation peaks in $l(r)$, except for MgCl$_2$, which is observed to induce 
small shift in peak positions to smaller separations. Most notably, however, the long range nature of $l(r)$ visibly disappears within 1.0M 
concentrations for all salts studied (FIG. 2(d)), fit results in TABLE I show that the long range component vanishes even at 0.15M). 
The long range behavior of $l(r)$ correlations in liquid water was shown to be predominantly governed by 
local fluctuations in the underlying hydrogen bond network of water~\cite{kanth10}. Observed effects of salts on $l(r)$ thus clearly indicate that the 
salts studied induce perturbations in the hydrogen bond network of water,- especially MgCl$_2$, for which the magnitude of effect indicates at 
highly non-local effects owing to the presence of strongly solvated Mg$^{2+}$ ions.

\subsection{Effect of salts on position - dipolar orientational correlations}
The position - dipole correlations have been computed using TIP4P-Ew water model alone. The oxygen - dipole correlation $d(r)$ for solute free liquid water is observed to 
display oscillatory solvation structure at small radial separations below $14\mathring{A}$ and vanish beyond the same (FIG. 3 (a)). The result is in good agreement 
with prior observations of the same correlation~\cite{kanth10}. Dissolved ions have strong electrostatic interactions with water molecules in their first few 
solvation shells. The solute - dipole correlations for ionic species are thus expected to deviate widely from the oxygen - dipole correlations in solute free water. Further, the nature of 
the correlations is expected to vary strongly between cationic and anionic species owing to their preferential interactions with the oxygen and hydrogen atoms of 
a water molecule respectively. The charge density of the ions can also be envisaged to play a critical role in the patterning of water molecules around ions. 
Thus, the solute - dipole correlations for ions are likely to be highly ion specific, and the same can have important consequences on the connectivity, as well as the 
dynamics of the hydrogen bond network both within and beyond the first few solvation shells. Rich literature exists on ion specific effects on the structure 
and dynamics of water molecules, encompassing both local and non-local effects~\cite{makbr2,tiel,tiel2,obr1,obr2,UBpre,yang,iru}. In the following, we investigate the effects of 
the presence of a single divalent (Mg$^{2+}$) or monovalent (Na$^+$) cation (with requisite Cl$^-$ anion(s) for charge neutrality) on the orientational behavior of water around them 
and compare the same with solute free water $d(r)$ correlations. To retain similar statistics for solute free water and salt - water systems, the $d(r)$ calculation for solute free 
water was carried out by considering a single \textit{tagged} water molecule at the center of the simulation box (since the same for salt - water systems can 
be carried out only over a single cation per frame). The results of the analyses are shown in FIG. 3(a,b). $d(r)$ for solute free water with 
TIP4P-Ew water model is observed to reproduce all qualitative trends of the same, previously reported with TIP5P water model~\cite{kanth10}.

\indent
As can be clearly seen from FIG. 3 (b), the presence of a cation has a significant effect on the orientation of water dipoles, causing them to align along the 
radial vector. The effect is strongest for the first solvation shell waters and gradually falls off with distance, with solvation peak structure. The correlation 
is long range, being non-vanishing at $>14\mathring{A}$, at least when single salt molecules are present in a box of water. Interestingly, solute - dipole $d(r)$ for ionic solutes falls 
off faster than the charge oscillations within spheres of increasing radii around ions (FIG. SIII of \textit{Supp. Info}~\cite{supplement}), indicating that the 
neighboring water molecules (hence hydrogen bonding) also have considerable effects on the structuring of water molecules around ions. The divalent cation (Mg$^{2+}$) 
has a stronger effect compared to the monovalent one (Na$^+$), while both result in $\sim$2 orders of magnitude enhancement over $d(r)$ correlations in solute free water. Further, 
anticorrelated regions present for solute free water are absent when observed from ion sites. 

\indent
Such preferential patterning of water molecules' dipoles, or equivalently of hydrogen bonding arms, around each ion can have important consequences on the long range 
dipolar correlations. As a consequence of the patterning, orientation fluctuations of neighboring water molecules are restricted as compared to those in solute free water. Such 
local perturbations to the underlying dynamics of hydrogen bond network is coupled to the long range orientational correlations, and more generally to the orientational 
entropy of the liquid, through the sum rule~\cite{kanth12,kanth13}. The ions can thus be envisaged as de-correlating centers, which along with their regions of influence, 
effectively screen bulk like water molecules from one another,- thus causing a decrease in the range of (ensemble averaged) long range correlations. The spatial extent of 
region of influence, as well as the strength of the patterning are critically related to the charge density of the ions~\cite{iru}. Thus, strongly solvated ions can be expected to 
affect the long range orientational correlations to a greater extent than weakly solvated ones. With increasing salt concentration, there is likely to be a two-fold effect. 
Trivially, there is an enhanced fraction of water in the orientationally restricted regions of influence, whose orientational fluctuations are suppressed. Further, an enhancement 
in the number of such de-correlation centers with increasing number of ions can be expected to result in a more effective screening. Our observations are generally 
consistent with such effects, with the exception of enhancement in $t(r)$ at lower salt concentrations. 

Interestingly, the decay of the solute - dipole correlation around both Na$^+$ and Mg$^{2+}$ ions show similar long range behavior. Fitting the $d(r)$ correlations 
for them for $r \geq10 \mathring{A}$ with the fit function defined in equation 5 ($a_2$ = 0), yields correlation lengths of (9.45$\pm$0.42) for Na$^+$ and (8.81$\pm$0.28) for Mg$^{2+}$. 
This indicates that, while the effects of ions on the orientations of first few solvation shell waters is strongly governed by the valency and charge density of the ions, the 
approach to pure water-like orientation can be independent of such parameters.

\subsection{Comparison of effects : hydrophilic and hydrophobic solute}
At the very low concentration of solutes studied, with only one solute molecule among $\sim$4100 water molecules, the effects of the solutes on the dipolar correlations 
$t(r)$ and $l(r)$ are expected to be extremely small. As shown in the inset plots of FIG. 3 (c,d), the correlations only differ marginally from that in absence of solutes~\cite{foot}. 
However, the long range parts of the correlations $t(r)$ and $l(r)$ can still be seen to be capable of distinguishing the effects 
of individual solute molecules, further emphasizing the applicability of the correlations in investigating orientational correlations among water molecules (FIG. 3 (c,d)). The $t(r)$ 
correlations for NaCl and MgCl$_2$ can be seen to have the same qualitative differences as observed at 0.15M concentration, both among themselves and with water in absence of solutes. 
$t(r)$ correlation for CH$_4$ is observed to be almost identical to that for solute free water, with indications of enhanced structuring, especially at larger separations. $l(r)$ correlation 
for CH$_4$, however shows clear differences, being enhanced compared to solute free water. $l(r)$ correlation can be observed to be reduced for MgCl$_2$, and comparable for NaCl when 
compared to solute free water. The results clearly show that small hydrophilic and hydrophobic solutes have contrasting effects on the long range orientational correlations among water molecules,- 
the former causing reduction and the latter leading to enhancement. $d(r)$ correlation around CH$_4$ shows no long range preferential orientation, as expected from its charge neutrality and 
size.

\section{Discussion}
\indent
The results in this study indicate that the presence of dissolved salts (ions) leads to considerable changes in the dipolar correlations. 
Such changes are both salt species and concentration dependent. The effects on the transverse and longitudinal parts of the 
dipolar correlations are considerably varied. The transverse part of the correlation is observed to be enhanced at lower (0.15M) 
concentrations for all salts with CsCl, NaCl and KCl leading to 
enhancement of statistically comparable magnitude. The presence MgCl$_2$ at 0.15M concentration is observed to lead to a greater 
enhancement in the transverse part for smaller separations, which is compensated at larger separations in comparison to the 
other salts. At higher salt concentration of 1.0M, the presence of MgCl$_2$ leads to strong reduction in the transverse 
alignment of water dipoles resulting in anticorrelations over shorter ($\sim12\mathring{A}$) separations, while the other salts 
studied retain the enhancement in correlation over solute free water. The longitudinal part of the dipolar correlation shows 
uniform reduction with salt concentration for all salts studied. The magnitude of reduction is further observed to be consistent 
with the relative positions of the cations in the Hofmeister cationic series for both salt concentrations studied. Interestingly, 
the long range exponential decay of the longitudinal correlation observed for solute free water is greatly reduced in the presence of 
salts, completely vanishing at the higher concentration studied. The reduction in the long range longitudinal correlation in 
presence of ions can be explained by considering ion locations as de-correlation centers. Within their regions of influence, ions 
result in orientational stratification of water molecules and a consequent restriction in their orientational fluctuations. 
Such local perturbations are coupled to the long range dynamics of the underlying hydrogen bond network through the sum rule. As 
a result, long range orientational correlations, whose molecular origin lies in the inherent fluctuations of the hydrogen bond 
network, get suppressed. 

\indent
Further, the varied responses of the transverse and longitudinal parts of the dipolar correlations to the presence of salts at lower 
concentration (0.15M), as well as the concentration dependent enhancement or reduction of the transverse part for the presence of the 
same salt (MgCl$_2$) further bolsters the criticism of the prevalent classification of ions along the lines of structure making and 
breaking.

\indent
The contrasting effects of hydrophilic and hydrophobic solutes on long range $l(r)$ correlations can have interesting implications in 
understanding the long range ($10 - 100\,\mathring{A}$) component of hydrophobic force. It was suggested in our earlier publication 
that $l(r)$ correlations can lead to a shape dependent attraction between two hydrophobic surfaces at large distances of 
separation~\cite{kanth10}, decaying with a correlation length of $\sim 12\mathring{A}$. Experimental evidences have shown that the 
force between hydrophobic surfaces, acting in spatial separation between ($10 - 100\,\mathring{A}$), is exponential in nature, 
and the range, as well as the magnitude of the same is reduced in presence of ionic impurities~\cite{meyer}. Our observations of 
reduction in the range of $l(r)$ correlations is consistent with the same. The observations for methane can not be extended to 
extended hydrophobic surfaces in a straightforward manner. However, sum rule for water dictates that long range $l(r)$ correlations 
in the presence of extended hydrophobic surfaces would be considerably altered compared to the same for water in absence of impurities, 
owing to the presence of dangling bonds at hydrophobic surfaces. Future research would be directed at studying the effects 
of larger hydrophobes as well as surfactants and osmolytes on the discussed correlations.

\section{Acknowledgments}
\indent
All the simulations in this work have been carried out on 1024-cpu Annapurna cluster at The Institute of Mathematical Sciences.


\begin{thebibliography}{53}
\bibitem[{foo()}]{foot}
\bibinfo{note}{In the presence of a single solute, various correlations d(r),
  l(r) and t(r) have no dependance on position and orientations of the solute
  molecules, as these have been averaged over all possible configurations.}
\end{thebibliography}
\end{document}